\documentclass[aps,showpacs,preprintnumbers,amsmath,amssymb,superscriptaddress,floatfix,a4paper,nofootinbib,11pt,floatfix,nofootinbib,onecolumn]{revtex4}
\usepackage{graphicx}
\usepackage{bm}
\usepackage{rotating}
\usepackage{array}
\usepackage{xcolor}
\usepackage{amsmath,amssymb}
\usepackage{mathrsfs}
\usepackage{graphicx}
\usepackage{color}
\usepackage{subfigure}
\usepackage{fancyhdr}
\usepackage{multirow}
\usepackage{float}
\usepackage{epsfig}
\usepackage{amsfonts}
\usepackage{bm}

\newcommand{\ba}{\begin{eqnarray}}
\newcommand{\ea}{\end{eqnarray}}

\begin{document}

\title{Thin-shell wormholes in de Rham-Gabadadze-Tolley massive gravity}

\author{Takol Tangphati} 
\email{ta_kol@hotmail.com}
\affiliation{Department of Physics, Faculty of Science, Chulalongkorn University, Bangkok 10330, Thailand}

\author{Auttakit Chatrabhuti} 
\email{auttakit@sc.chula.ac.th}
\affiliation{Department of Physics, Faculty of Science, Chulalongkorn University, Bangkok 10330, Thailand}

\author{Daris Samart} 
\email{dsamart82@gmail.com}
\affiliation{Department of Physics, Faculty of Science, Khon kaen University, 123 Mitraphap road, Khon Kaen, 40002, Thailand}

\author{Phongpichit Channuie} 
\email{channuie@gmail.com}
\affiliation{College of Graduate Studies, Walailak University, Thasala, Nakhon Si Thammarat, 80160, Thailand}
\affiliation{School of Science, Walailak University, Thasala, Nakhon Si Thammarat, 80160, Thailand}
\affiliation{Research Group in Applied, Computational and Theoretical Science (ACTS), Walailak University, Thasala, Nakhon Si Thammarat, 80160, Thailand}
\affiliation{Thailand Center of Excellence in Physics, Ministry of Education, Bangkok 10400, Thailand}

\date{\today}

\begin{abstract}

In this work, we study the thin-shell wormholes in dRGT massive gravity. In order to joint two bulks of the spacetime geometry, we first derive junction conditions of the dRGT spacetime. This results the dynamics of the spherical thin-shell wormholes in the dRGT theory. We show that the massive graviton correction term of the dRGT theory in the Einstein equation is represented in terms of the effective anisotropic pressure fluid. However, if there is only this correction term, without invoking exotic fluids, we find that the thin-shell wormholes can not be stabilized. We then examine the stability conditions of the wormholes by introducing four existing models of the exotic fluids at the throat. In addition, we analyze the energy conditions for the thin-shell wormholes in the dRGT massive gravity by checking the null, weak, and strong conditions at the wormhole throat. We show that in general the classical energy conditions are violated by introducing all existing models of the exotic fluids. Moreover, we quantify the wormhole geometry by using the embedding diagrams to represent a thin-shell wormhole in the dRGT massive gravity.

\end{abstract}

%\pacs{Valid PACS appear here}

\maketitle

%%%%%%%%%%%%%%%%%%%%%%%%%%%%%%%%%%%%%%%%
%%%%%%%%%%%%%%%%%%%%%%%%%%%%%%%%%%%%%%%
%%%%%%%%%%%%%%%%%%%%%%%%%%
\section{Introduction}
%%%%%%%%%%%%%%%%%%%%%%%%%%
General theory of relativity provides an elegant mathematical description of the spacetime geometry and matter described by the energy-momentum tensor. One of the viable solutions to Einstein’s field equations provides a static and spherically symmetric black hole \cite{Schwarzschild:1916uq}. The recent detection of gravitational waves (GWs) \cite{Abbott:2016blz} demonstrated that stellar-mass black holes really exist in Nature. Interestingly, Ludwing Flamm \cite{Flamm:1916} realized in 1916 that Einstein’s equations allowed for another solution currently known as a white hole. However, in contrast to black holes, it was believed that white holes eject matter and light from their event horizon. These two solutions could represent two different regions in spacetime connected by a conduit. This conduit was named later a “bridge” and in 1935, Einstein and Rosen used the theory of general relativity to propose the existence of ”bridges” through space-time \cite{Einstein:1935tc}. Historically some decades later, Misner and Wheeler first introduced the term “wormhole” in Ref.\cite{Misner:1957mt}. The paper came to
focus of researchers and caused new studies over structural characteristics of wormholes. However, the original version of wormholes was later ruled out because they are not traversable. This means that its throat opens and closes so quickly. 

However, in order to prevent the wormhole’s throat from closing, one can add a scalar field coupled to gravity. These new classes of solutions provide a more general class of wormholes firstly proposed by Ellis \cite{Ellis:1973yv} and independently by Bronnikov \cite{Bronnikov:1973fh}. The main problem of wormholes is that they are supported by exotic matter; a kind of matter which violates known energy conditions. Some conditions introduced by Morris and Throne in 1988 for the wormholes to be traversable can be found in Ref.\cite{Morris:1988cz}. These solutions are obtained by considering an unusual type of matter which can maintain the structure of the wormhole. In addition, this exotic matter with negative energy density satisfies the flare-out condition  and  violates  weak  energy  condition \cite{Morris:1988cz,Morris:1988tu}.

Alternative theories of gravity give opportunities for existence of traversable wormholes. In the literature, many authors have intensively studied  various  aspects  of  traversable wormhole (TW) geometries  within  different  modified  gravitational  theories \cite{Harko:2013aya,Lobo:2009ip,Bohmer:2011si,Zangeneh:2015jda,Clement:1983fe,Bronnikov:2010tt,Jusufi:2017drg,Jusufi:2016leh,Ovgun:2018xys,Shaikh:2018yku,Shaikh:2016dpl,Ovgun:2018prw,MontelongoGarcia:2010xd,Rahaman:2016jds,Rahaman:2014dpa,Jamil:2013tva,Rahaman:2012pg}. Some of these include $f(R)$ and $f(T)$ theories, see e.g., \cite{Bahamonde:2016jqq,Bahamonde:2016ixz,Jamil:2012ti,Jamil:2008wu}. In addition, it turns out that the effect of phantom energy can evoke a natural existence of traversable wormholes \cite{Lobo:2005us}. Moreover, the Einstein-dilaton-Gauss-Bonnet gravity allows for a traversable wormhole solution studied by Ref.\cite{Kanti2011}. More recently, it is proposed that Casimir energy is introduced as a potential source to generate a traversable wormhole \cite{Garattini:2019ivd}. There have been some particular thin-shell wormholes constructed from black holes, see e,g, \cite{Halilsoy:2013iza,Ovgun:2016ijz,Richarte:2017iit}.

In the present work, we study the thin-shell wormholes in dRGT massive gravity. The study of massive gravity has begun prior to the discovery of the accelerated expansion of the universe. In 1939, Fierz and Pauli used a linear theory of massive gravity as a mass term of the graviton \cite{Fierz:1939ix}. Unfortunately, there are some flaws of the proposed model \cite{Zakharov:1970cc,vanDam:1970vg} where the asymptotic-massless limits of the linear theory do not satisfy the GR prediction. Later on, the author of Ref.\cite{Vainshtein:1972sx} suggested that the non-linear approach in the massive gravity theory might be able solve the problem. However, it inevitably leads to a new problem called the Boulware-Deser (BD) ghost  \cite{Hinterbichler:2011tt}. However, in 2010, the BD ghost was completely eliminated by the new non-linear version of massive gravity proposed by de Rham, Gabadadze, and Tolley (dRGT) \cite{deRham:2010kj}. Since then the dRGT was targeted as one of the compelling scenarios when studying the universe in the cosmic scale. Additionally, there have been many interesting articles regarding the applications of the dRGT massive gravity to the exotic objects, e.g., black holes \cite{Ghosh:2015cva, Chabab:2019mlu, Boonserm:2017qcq}. 

In this work, we study the thin-shell wormholes in dRGT massive gravity. We first take a short review of the dRGT model of the nonlinear massive gravity in the Sec.\ref{dRGT_model}. In Sec.\ref{approach_dRGT_wormholes}, we consider the mathematical setup in order to study the thin-shell wormhole. We study junction conditions allowing to glue two identical dRGT spacetimes. In addition, we study stability analyses of the dRGT thin-shell wormhole by considering four existing exotic fluid models in Sec,\ref{stability}. We also check the null, weak, and strong conditions at the wormhole throat for all models present in Sec.\ref{ener}. Moreover, we quantify the wormhole geometry by using the embedding diagrams to represent a thin-shell wormhole in the dRGT massive gravity in Sec.\ref{dia}. Finally, we discuss our main results and conclude our findings in the last section. In this work, we use the geometrical unit such that $G=1$.

%%%%%%%%%%%%%%%%%%%%%%%%%%%%%%%%%%%%%%%%%%%
\section{A short recap of massive gravity} \label{dRGT_model}
%%%%%%%%%%%%%%%%%%%%%%%%%%%%%%%%%%%%%%%%%%
In this section, we assume that the universe is undergoing an accelerating phase, according to the modified massive gravity theory called dRGT model \cite{deRham:2010kj,deRham:2010ik}. The solutions of the model will be derived shortly. We begin with the action representing the dRGT model on the manifold ${\cal M}$ given by
\begin{equation}
S = \int_M d^4 x \sqrt{-g} \frac{1}{16\pi G}\big( R + m^2_g \mathcal{U}(g,\phi^a) \big),
\label{action_dRGT}
\end{equation}
where $\sqrt{-g}$ is the volume element in 4-dimensional manifold $\mathcal{M}$ and the potential $\mathcal{U}$ is defined by 
\begin{eqnarray}
\mathcal{U} = \mathcal{U}_2 + \alpha_3 \mathcal{U}_3 + \alpha_4 \mathcal{U}_4\,,
\end{eqnarray} 
where $\mathcal{U}_{2},\,\mathcal{U}_{3}$ and $\mathcal{U}_{4}$ are given by
\begin{eqnarray}
\mathcal{U}_2 &=& [\mathcal{K}]^2 - [\mathcal{K}^2], \nonumber \\
\mathcal{U}_3 &=& [\mathcal{K}]^3 - 3[\mathcal{K}][\mathcal{K}^2] + 2 [\mathcal{K}^3], \nonumber \\
\mathcal{U}_4 &=& [\mathcal{K}]^4 - 6[\mathcal{K}]^2[\mathcal{K}^2] + 8[\mathcal{K}][\mathcal{K}^3] + 3[\mathcal{K}^2]^2 - 6[\mathcal{K}^4], \nonumber \\
\mathcal{K}^{\mu}_{\nu} &=& \delta^{\mu}_{\nu} - \sqrt{g^{\mu\lambda} \mathcal{F}_{ab} \partial_{\lambda}\phi^a \partial_{\nu}\phi^b }.
\label{kappa_dRGT}
\end{eqnarray}
Here a bracket $[\quad]$ represents the trace of the tensor and $\mathcal{F}_{ab}$ is the fiducial metric which is chosen as 
\begin{equation}
\mathcal{F}_{ab} = 
\begin{pmatrix}
0 & 0 & 0 & 0 \\
0 & 0 & 0 & 0 \\
0 & 0 & k^2 & 0 \\
0 & 0 & 0 & k^2 \text{sin}^2 \theta \\
\end{pmatrix},
\label{fd_matrix}
\end{equation}
where $k$ is a positive constant and the unitary gauge is used as 
\begin{equation}
	\phi^a = x^{\mu} \delta^a_{\mu}\,.
\end{equation}
In addition, the parameters $\alpha_{3,4}$ are the parameters of the dRGT theory and we will relate these parameters with the graviton mass in the latter.

Varying the gravitational action in Eq.(\ref{action_dRGT}) with respect to the metric, $g^{\mu\nu}$, the Einstein equation of the dRGT massive gravity is given by,
\begin{equation}
G_{\mu\nu} + m^2_g X_{\mu\nu} = 0,
\label{Einstein_dRGT}
\end{equation}
where $X_{\mu\nu}$ is defined by
\begin{eqnarray}
X_{\mu\nu} &=& \frac{1}{\sqrt{-g}}\frac{\delta \sqrt{-g} \mathcal{U}}{ \delta g^{\mu\nu}}
\nonumber\\
&=& \mathcal{K}_{\mu\nu} -\alpha \left[ \big(\mathcal{K}^2 \big)_{\mu\nu} - [\mathcal{K}]\mathcal{K}_{\mu\nu} + \frac12 g_{\mu\nu}\big( [\mathcal{K}]^2 - [\mathcal{K}^2]\big)\right]
\label{X-munu}\\
&& +\, 3\beta\left[ \big(\mathcal{K}^3\big)_{\mu\nu} - [\mathcal{K}]\big( \mathcal{K}^2\big)_{\mu\nu} + \frac12 \mathcal{K}_{\mu\nu}\big( [\mathcal{K}]^2 - [\mathcal{K}^2]\big) -\frac16 g_{\mu\nu}\big( [\mathcal{K}]^3 - 3[\mathcal{K}][\mathcal{K}^2] + 2[\mathcal{K}^3]\big)\right] ,
\nonumber
\end{eqnarray}
where the parameters $\alpha$ and $\beta$ are related to $\alpha_{3,4}$ from the action in Eq.(\ref{action_dRGT}) via
\begin{eqnarray}
\alpha = 1 + 3\alpha_3\,,\qquad \beta = \alpha_3 + 4\alpha_4.
\end{eqnarray}
%Also note that, the surface term in the variational method is omitted since $\delta g_{\mu \nu}$ on the hypersurface of the manifold $\partial M$ is zero.
%\begin{equation}
%	\delta g_{\mu\nu}|_{\partial M} = 0
%\end{equation}
In addition, one finds that the Einstein equation of the dRGT massive gravity is reduced to standard GR in the $m_g \to 0$ limit.
\begin{figure}[h]	
	\includegraphics[width=10cm]{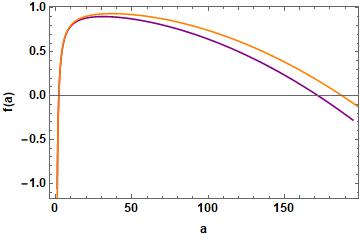}
	\centering
	\caption{A figure shows the behaviors of $f(r)$ against $r$. For simplicity, we have used $M=1.0,\,G=1.0,\,\zeta=0.0$. We labelled lines with orange and purple colors for $\Lambda=0.0001,\,\gamma=0.001$ and $\Lambda=0.0001,\,\gamma =0.00002$, respectively.}
	\label{frplot}
\end{figure}
Having used the static spherical symmetric spacetime, we obtain the explicit form of the line element of the dRGT massive gravity which reads \cite{deRham:2010kj, deRham:2010ik}
\begin{equation}
ds^2 = g_{\mu\nu} dx^{\mu}dx^{\nu} = - f(r) dt^2 + \frac{dr^2}{f(r)} + r^2 d\theta^2 + r^2 \text{sin}^2\theta d\phi^2,
\label{metric_dRGT_spherical}
\end{equation}
where the function $f(r)$ can be written as
\begin{eqnarray}
f(r) = 1 - \frac{2M}{r} - \frac{\Lambda r^2}{3} + \gamma r + \zeta,
\label{fr_dRGT}
\end{eqnarray}
where $M$ is the mass parameter, $\Lambda$ is the effective cosmological constant, and $\gamma$ and $\zeta$ are new parameters and they are linear combinations of the parameters in the dRGT massive gravity via the following relations,
\begin{eqnarray}
\Lambda \equiv -3m_g^2(1+ \alpha + \beta), \quad \gamma \equiv -m_g^2k(1 + 2\alpha + 3\beta),\quad \zeta \equiv m_g^2k^2(\alpha + 3\beta). 
\label{para-fr}
\end{eqnarray}
%
%The non-zero components of $G_{\mu\nu}$ are
%\begin{eqnarray}
%G_{tt} &=& -\frac{f(r)(-1 + f(r) + r f'(r) )}{r^2} \label{Gtt}  \\
%G_{rr} &=& \frac{(-1 + f(r) + r f'(r) )}{f(r) r^2} \label{Grr} \\	
%G_{\theta \theta} &=& \frac{1}{2} r (2f'(r) + r f''(r) ) \label{Gthth}\\
%G_{\phi \phi} &=& \frac{1}{2} r \: \text{sin}^2 \theta (2f'(r) + r f''(r) ) %\label{Gphiphi}
%\end{eqnarray}
The behaviors of how a function $f(r)$ depends on $r$ is displayed in Fig.(\ref{frplot}). According to the Einstein equation of the dRGT massive gravity in Eq.(\ref{Einstein_dRGT}), one might identify the $m_g^2X_{\mu\nu}$ term as the effective energy-momentum tensor. By using the explicit form of the metric tensors in Eq. (\ref{fr_dRGT}), we directly compute components of the $m_g^2X_{\mu\nu}$ in Eq.(\ref{X-munu}). The components of the $m_g^2 X_{\nu}^\mu$ are given by \cite{Burikham:2016cwz,Kareeso:2018xum,Panpanich:2018cxo}
\begin{eqnarray}
\rho_g(r) &\equiv& -\frac{m_g^2}{8\pi G}X_t^t 
= -\frac{m_g^2}{8\pi G}\left( \frac{3r -2k}{r} + \frac{\alpha(3r-k)(r-k)}{r^2} + \frac{3\beta(r-k)^2}{r^2}\right),
\label{rho-g}\\
p_g^{(r)}(r) &\equiv& -\frac{m_g^2}{8\pi G}X_r^r = \frac{m_g^2}{8\pi G}\left( \frac{3r -2k}{r} + \frac{\alpha(3r-k)(r-k)}{r^2} + \frac{3\beta(r-k)^2}{r^2}\right),
\label{p-g-r}\\
p_g^{(\theta,\phi)}(r) &\equiv& -\frac{m_g^2}{8\pi G}X_{\theta,\phi}^{\theta,\phi} 
= \frac{m_g^2}{8\pi G}\left( \frac{3r -k}{r} + \frac{\alpha(3r-2k)}{r} + \frac{3\beta(r-k)}{r}\right).
\label{p-g-theta}
\end{eqnarray}
With help of Eq.(\ref{para-fr}) the expressions of the $k$, $\alpha$ and $\beta$ are re-written in terms of $\Lambda$, $\gamma$ and $\zeta$ by
\begin{eqnarray}
k &=& \frac{\gamma+\sqrt{\gamma^2 + (m_g^2+\Lambda)\zeta}}{m_g^2+\Lambda}, \nonumber\\
\alpha &=& -\frac{\gamma^2 + (2 m_g^2 +\Lambda)\zeta-\gamma\sqrt{\gamma^2 + (m_g^2 + \Lambda)\zeta}}{m_g^2\zeta}, \nonumber\\
\beta &=& \frac{2 \Lambda}{3 m_g^2} + \frac{\gamma^2 + m_g^2\zeta -\gamma\sqrt{\gamma^2 + (m_g^2 + \Lambda)\zeta}}{m_g^2\zeta}.
\label{para-efe}
\end{eqnarray}
In addition, we found that the parameters $k$, $\alpha$ and $\beta$ are finite values in the $\zeta\to 0$ limit i.e.,
\begin{eqnarray}
k = \frac{2\gamma}{m_g^2+\Lambda},\qquad \alpha = -3 \beta = -\frac{3}{2}-\frac{\Lambda}{2 m_g^2}. \nonumber
\end{eqnarray}
In this limit, above relation gives a well define for tuning the positive value and the smallness of the effective cosmological constant, $\Lambda$, with the observed values from the cosmological data. More importantly, it has been shown that the effective energy-momentum tensor, $X_{\mu\nu}$ exhibits its behavior like anisotropic dark energy fluid i.e., $p_g^{(r)} = -\rho_g$ see more detail discussions and applications in Refs. \cite{Burikham:2016cwz,Kareeso:2018xum,Panpanich:2018cxo}. 

We close this section with the inclusion of the matter fluid to the Einstein equation. By using the standard method in GR, the Einstein equation with the (fluid) matter field of the dRGT massive gravity is given by
\begin{eqnarray}
G_{\mu\nu} &=& 8\pi G T_{\mu\nu},\qquad T_{\mu\nu} = T_{\mu\nu}^{(f)} + T_{\mu\nu}^{(g)}, \qquad 
\\
T_{\mu\nu}^{(f)} &=& (\sigma + p) u_\mu u_\nu + p g_{\mu\nu}, \qquad
\nonumber\\
T_{\mu\nu}^{(g)} &=& -\frac{m_g^2}{8\pi G}X_{\mu\nu} = \Big( \rho_g + p_g^{(\perp)}\Big)u_\mu u_\nu + p_g^{(\perp)} g_{\mu\nu} + \Big( p_g^{(r)} - p_g^{(\perp)}\Big)\chi_\mu \chi_\nu,
\label{Einstein_dRGT_matter}
\end{eqnarray}
where $\sigma$ and $p$ are the energy density and isotropic pressure of the fluid matter while the components of the effective energy momentum tensor of the dRGT massive gravity $\big(T_{\mu\nu}^{(g)}\big)$ have been displayed in Eqs.(\ref{rho-g}, \ref{p-g-r}, \ref{p-g-theta}). The $u_\mu$ is timelike unit vector, the $\chi_\mu$ is the spacelike unit vector orthogonal to the $u_\mu$ and the angular plane and $p_g^{(\perp)} = p_g^{(\theta)} = p_g^{(\phi)}$. The total energy momentum tensor might re-write in the compact matrices as,
\begin{eqnarray}
T_\nu^\mu =T_{\nu}^{(f),\,\mu} + T_{\nu}^{(g),\,\mu} 
= \left(
\begin{array}{cccc}
-\sigma-\rho_g~ \quad 0\quad~ \,~\quad 0\quad~\,~\quad 0\quad~\\
~ \quad 0\quad~ \,\,p + p_g^{(r)}~ \quad 0\quad~ \,~ \quad 0\quad~ \\
~\,\, \quad 0\quad~ \,~ \quad 0\quad~ \,p + p_g^{(\perp)}~ \quad 0\quad~ \\
~\,\,\, \quad 0\quad~ \,~ \quad 0\quad~ \,~ \quad 0\quad~ \,p + p_g^{(\perp)}
\end{array}
\right).
\end{eqnarray}
Moreover, the conservation of the total energy momentum tensor is hold as well as for the perfect fluid and the massive graviton parts as,
\begin{eqnarray}
\nabla^\mu T_{\mu\nu} = 0, \qquad \nabla^\mu T_{\mu\nu}^{(f)} = 0, \qquad \nabla^\mu T_{\mu\nu}^{(g)} = 0.
\end{eqnarray}

%%%%%%%%%%%%%%%%%%%%%%%%%%%%%%%%%%%%%%%%
\section{The thin-shell wormholes in dRGT spacetimes} \label{approach_dRGT_wormholes}
%%%%%%%%%%%%%%%%%%%%%%%%%%%%%%%%%%%%%%%%
\subsection{Junction conditions in dRGT theory}
In this work, the wormhole helps joining two different dRGT spacetimes. It behaves like a surface between two bulks and it is called thin-shell \cite{Visser:1995cc}. In order to join two different manifolds, we will construct a thin-shell wormhole of the dRGT massive gravity following the standard approach proposed in Refs.\cite{Poisson:2009pwt,Ovgun:2016qiu,Aviles:2019xae}. Consider two distinct spacetime manifolds, ${\cal M}_{+}$ and ${\cal M}_{-}$, with metrics given by $g^{+}_{\mu\nu}(x^{\mu}_{+})$ and $g^{-}_{\mu\nu}(x^{\mu}_{-})$, in terms
of independently defined coordinate systems $x^{\mu}_{+}$ and $x^{\mu}_{-}$. The manifolds are bounded by hypersurfaces $\Sigma_{+}$ and $\Sigma_{-}$, respectively, with induced metrics $h^{+}_{ab}$ and $h^{-}_{ab}$. Regarding the Darmois-Israel formalism, the coordinates on ${\cal M}$ can be choosen as $x^{\mu}=(t, r, \theta, \phi)$, while for the coordinates on the induced metric $\Sigma$, we write $y^{a} = (\tau, \theta, \phi)$ being the intrinsic coordinates. Note that the hypersurfaces are isometric, i.e., $h^{+}_{ab}(y^{a})=h^{-}_{ab}(y^{a})=h_{ab}(y^{a})$. A single manifold ${\cal M}$ is obtained by gluing together ${\cal M}_{+}$
and ${\cal M}_{-}$ at their boundaries, i.e., ${\cal M}={\cal M}_{+}\cup{\cal M}_{-}$, with the natural identification of the boundaries $\Sigma_{+}=\Sigma_{-}=\Sigma$.

We describe the two different manifolds as follows \cite{Poisson:2009pwt, Ovgun:2016qiu}:
\begin{equation}
\mathcal{M}_{\pm} = \{ x^{\mu}_{\pm}\,\,|\, t_\pm \geq T_\pm(\tau) \text{ and } r \geq a(\tau) \},
\label{Manifolds_up_down}
\end{equation}
where the plus (minus) sign means the upper (the lower) spacetime. The line elements of the manifolds are given by
\begin{equation}
	ds_\pm^2 = g_{\mu\nu}^\pm dx^{\mu}dx^{\nu} = - f_\pm(r) dt^2 + \frac{dr^2}{f_\pm(r)} + r^2 d\theta^2 + r^2 \text{sin}^2\theta d\phi^2.
	\label{metric_dRGT_bulk}
\end{equation}
Both different manifolds are linked by the (co-moving) thin-shell and the hypersurface $\Sigma$ is parametrized by \cite{Poisson:2009pwt, Ovgun:2016qiu, Kokubu:2017, Kim:2018dnl}
\begin{equation}
\Sigma = \{ y^{a}\,\,|\, t_\pm = T_\pm(\tau) \text{ and } r = a(\tau) \}.
\label{Hypersur_thin_shell}
\end{equation}
Thus the line element of the thin-shell reads
\begin{eqnarray}
ds_{\Sigma}^2 &=& g_{\alpha\beta}dx^{\alpha}dx^{\beta} =  g_{\alpha\beta} \bigg( \frac{\partial x ^ {\alpha}}{\partial y^a} dy^a \bigg) \bigg( \frac{\partial x ^ {\beta}}{\partial y^b} dy^b \bigg) = h_{ab}dy^a dy^b \nonumber \\
&=& -d\tau^2 + a^2(\tau) d\Omega^2, \qquad d\Omega^2 = d\theta^2 + \text{sin}^2\theta d\phi^2,
\label{metric_Thin_dhell}
\end{eqnarray}
where $y^a = y^a(x^{\mu})$ is a coordinate on the hypersurface $\Sigma$ and $h_{ab}$ is called the induced metric or the first fundamental form on the hypersurface $\Sigma$, \cite{Poisson:2009pwt, Ovgun:2016qiu, Kokubu:2017}
\begin{equation}
h_{ab} \equiv g_{\alpha\beta} e^{\alpha}_{a} e^{\beta}_{b}.
\end{equation}
It is worth mentioning here that the throat must satisfy the Israel junction conditions and these provide the following coordinate choice
\begin{eqnarray}
-f_\pm(a)\dot T_\pm^2 + \frac{\dot a^2}{f_\pm(a)} = 1,\qquad \forall\tau, 
\end{eqnarray}
where dots denote derivatives respect to a conformal time, i.e., ``$\;\dot\;\equiv d/d\tau\,$". The induced metric is a tangent component of $g_{\alpha \beta}$ on the hypersurface $\Sigma$. Then, the normal vector component $n_{\alpha}$ of the metric tensor on the hypersurface is defined as follows \cite{Poisson:2009pwt, Ovgun:2016qiu, Kokubu:2017}:
\begin{equation}
n_{\alpha} \equiv \frac{F(r, a(\tau))_{, \alpha}}{|F(r, a(\tau))_{, \beta} F(r, a(\tau))^{, \beta}|^{1/2}},
\label{normal_vector_hypersurface}
\end{equation}
where $F(r, a(\tau)) \equiv r - a(\tau) = 0$ is the hypersurface function and $a(\tau)$ is the throat radius of the thin-shell wormhole. Note that the Greek indices ($x^{\alpha}, x^{\beta}, \dots$) are the coordinates on Manifold $M_{\pm}$ while the Latin indices ($x^{a}, x^{b}, \dots$) are the coordinates on the hypersurface $\Sigma$. The metric tensor on the hypersurface can be split into two parts \cite{Poisson:2009pwt, Ovgun:2016qiu, Kokubu:2017} as
\begin{equation}
g_{ab} = h_{ab} + \epsilon n_a n_b,
\label{gab_h_nn}
\end{equation}
where $\epsilon$ represent the types of thin-shell with $\epsilon = -1,0,+1$ being the space-like, null-like and time-like, respectively. %Here in this work, we consider the time-like thin-shell in the latter study.

In the next step, we are going to derive the junction conditions on the hypersurface between the boundaries of two different manifolds ($\partial\mathcal{M}$). It is well known that in order to connect two manifolds one needs to add the action of the boundary terms or the Gibbons-Hawkings terms into the total action. Then the total gravitational action of the dRGT massive gravity with the matter fields is given by \cite{Aviles:2019xae, Padilla:2012ze}
\begin{eqnarray}
S_{\rm total} &=& \int_{\mathcal{M}_+} d^4 x \sqrt{-g^+}\left(\frac{1}{16\pi G} \big( R^+ + m^2_g \mathcal{U}(g^+,\phi^a) \big) + \mathcal{L}_{\rm matter}^+\right) 
\nonumber\\
&+& \frac{1}{8\pi G} \int_{\partial\mathcal{M}_+} d^3 y \sqrt{-h^+} K^+ 
\nonumber\\
&+& \int_{\mathcal{M}_-} d^4 x \sqrt{-g^-}\left(\frac{1}{16\pi G} \big( R^- + m^2_g \mathcal{U}(g^-,\phi^a) \big) + \mathcal{L}_{\rm matter}^-\right) 
\nonumber\\
&+& \frac{1}{8\pi G} \int_{\partial\mathcal{M}_-} d^3 y \sqrt{-h^-} K^-
\nonumber\\
&+& \int_{\Sigma} d^3y \sqrt{-h} \,\mathcal{L}_{\rm matter}^{\Sigma}
\label{total-action}
\end{eqnarray}
where $\sqrt{-h}$ is the volume element on the 3-dimensional hypersurface and $K$ is the trace of the extrinsic curvature $K_{ab}$ on the thin-shell with $K \equiv K^a_a = h_{ab} K^b_a$. Here the $\mathcal{L}_{\rm matter}^{\Sigma} = \mathcal{L}_{f}^{\Sigma} + \mathcal{L}_{g}^{\Sigma}$ is composed of the two types of fluids (perfect fluid, $\mathcal{L}_{f}^{\Sigma}$ and massive gravity fluid, $\mathcal{L}_{f}^{\Sigma}$) which are localized on the hypersurface. The extrinsic curvature can be calculated via the following equation
\begin{eqnarray}
K_{ab} &=& -n_{\alpha} \bigg[ \frac{d^2  x^{\alpha} }{dy^a dy^b} + \Gamma^{\alpha}_{\beta \gamma} \frac{dx^{\beta}}{dy^{a}} \frac{dx^{\gamma}}{dy^{b}} \bigg].
\label{extrinsic_curvature}
\end{eqnarray}
More importantly, it is worth to discuss about the dRGT massive gravity and the matter field on the hypersurface. Let us first discuss about the Lagrangian matter on the hypersurface, $\mathcal{L}_{\rm matter}^{\Sigma}$. It has been proven and demonstrated in Refs.\cite{Aviles:2019xae, Padilla:2012ze} for the scalar field matter case that the Lagrangian of the matter field on the hypersurface has the same form of the Lagrangian in the bulk but the metric tensor $g_{\mu\nu}$ in the bulk is replaced by the induced metric on the hypersurface, $h_{ab}$\,. 
%According to Refs.\cite{Aviles:2019xae, Padilla:2012ze}, on the other hand, the non-linear massive gravity potential, $U(h,\phi^a)$ on the hypersurface is constructed in the same manner by substituting $g_{\mu\nu} ~\to~ h_{ab}$ into the $U(g,\phi^a)$ function. The $U(h,\phi^a)$ and $F_{ab}$ on the hypersurface are written as
%\begin{eqnarray}
%U = U_2 + \alpha_3 U_3 + \alpha_4 U_4\,,
%\end{eqnarray} 
%where $U_{2},\,U_{3}$ and $U_{4}$ are given by
%\begin{eqnarray}
%U_2 &=& [\kappa]^2 - [\kappa^2], \nonumber \\
%U_3 &=& [\kappa]^3 - 3[\kappa][\kappa^2] + 2 [\kappa^3], \nonumber \\
%U_4 &=& [\kappa]^4 - 6[\kappa]^2[\kappa^2] + 8[\kappa][\kappa^3] + 3[\kappa^2]^2 - %6[\kappa^4], \nonumber \\
%\kappa^{\mu}_{\nu} &=& \delta^{\mu}_{\nu} - \sqrt{h^{\mu\lambda} F_{ab} %\partial_{\lambda}\phi^a \partial_{\nu}\phi^b }.
%\label{kappa_surface_dRGT}
%\end{eqnarray}
%In this work, the induced fiducial metric on the hypersurface is given by
%\begin{equation}
%F_{ab} = 
%\begin{pmatrix}
 % 0 & 0 & 0 \\
 %0 & k^2 & 0 \\
 %0 & 0 & k^2 \text{sin}^2 \theta \\
%\end{pmatrix}.
%\label{fd_induce_matrix}
%\end{equation}
In the following, we will apply the variational principle to the total gravitational action in Eq.(\ref{total-action}) to obtain the equation of motion for our study. Varying the total action, one finds,
\begin{eqnarray}
	\delta S &=&  \int_{\mathcal{M}_{+}} d^4x \sqrt{-g^+}\left( \frac{1}{16\pi G}\big( G_{\alpha \beta}^+ + m_g^2 X_{\alpha \beta}^+ \big) + T_{\alpha\beta}^{(f)\,,+}\right)\delta g^{ \alpha \beta}_+ 
	\nonumber \\
	&+& \int_{\partial\mathcal{M}^{+}} d^3 y \sqrt{-h^+} \frac{1}{8\pi G}\big( K_{ab}^+ - h_{a b}^+ K^+ \big) \delta h^{a b}_+ 
	\label{vary_action_complete}
	\nonumber \\
	&+& \int_{\mathcal{M}_{-}} d^4x \sqrt{-g} \left( \frac{1}{16\pi G}\big( G_{\alpha \beta}^- + m_g^2 X_{\alpha \beta}^- \big) + T_{\alpha\beta}^{(f)\,,-}\right) \delta g^{\alpha \beta}_- 
	\nonumber \\
	&+&  \int_{\partial\mathcal{M}_{-}} d^3 y \sqrt{-h^-}\frac{1}{8\pi G} \big( K_{ab}^- - h_{a b}^- K^- \big) \delta h^{a b}_- 
	\nonumber \\
	&+& \int_{\Sigma} d^3y \: \sqrt{-h}\, (-1)\big( t_{ab} + Y_{ab} \big) \delta h^{ab} \,.
\end{eqnarray}
Using the definition of the energy momentum tensor of the fluid on the thin-shell, $t_{b}^a$ takes the form
\begin{eqnarray}
	t_{b}^a &=& - \frac{1}{\sqrt{-h}} \frac{\delta}{\delta h_a^{b}} \Big( \sqrt{-h} \,\mathcal{L}_{f}^{\Sigma} \Big)
	= (\sigma + p)u^a u_b + p h_b^a.
	\label{fluid_thin_shell}
\end{eqnarray}
For the massive gravity fluid, the $Y_{b}^a$ tensor is written by
\begin{eqnarray}
	Y^a_b &=& - \frac{1}{\sqrt{-h}} \frac{\delta}{\delta h_a^{b}} \Big( \sqrt{-h} \,\mathcal{L}_{g}^{\Sigma} \Big)
	= (\rho_g + p_g^{(\perp)})u^a u_b + p_g^{(\perp)} h_b^a.
	\label{dRGT_thin_shell}
\end{eqnarray}
In order to find the equation of motion on the thin-shell wormhole, we consider the variation of the action on the boundaries $\partial\mathcal{M}_\pm$ and hypersurface $\Sigma$ with respect to (w.r.t.) the induced metric $h^{ab}$. After performing the variation, we find  
\begin{eqnarray}
	\frac{\delta S_{\rm total}}{\delta h^{ab}} 
	&=& \int_{\partial\mathcal{M}_{+}} d^3 y \sqrt{-h^+} \frac{1}{8\pi G}\big( K_{c d}^+ - h_{c d}^+ K^+ \big) \frac{\delta h^{c d}_+}{\delta h^{ab}} 
	\nonumber \\
	&+&  \int_{\partial\mathcal{M}_{-}} d^3 y \sqrt{-h^-} \frac{1}{8\pi G}\big( K_{c d}^- - h_{c d}^- K^- \big) \frac{\delta h^{c d}_-}{\delta h^{ab}} 
	\nonumber \\
	&-& \int_{\Sigma} d^3y \:  \sqrt{-h} \: \big( t_{ab} + Y_{ab} \big) .
	\label{vary_thin_shell}
\end{eqnarray}
The normal vector $n^a$ of the hypersurface points from $\mathcal{M}_-$ to $\mathcal{M}_+$. Then we can choose $n^a_- = n^a = -n^a_+$ in which the extrinsic curvature on each side is related via \cite{Aviles:2019xae,Padilla:2012ze}
\begin{equation}
	K^+_{ab}(n^a_+) = - K^+_{ab}(n^a),
\end{equation}
and
\begin{equation}
	K^-_{ab}(n^a_-) = K^-_{ab}(n^a).
\end{equation}
Moreover, the induced metric $h_{ab}$ are the same on both sides of the boundaries, i.e., $h_{ab}^+ = h_{ab} = h_{ab}^-$. By varying the total action on the hypersurface w.r.t. the induced metric, we finally obtain
\begin{eqnarray}
	\frac{\delta S_{\text{total}}}{\delta h^{ab}} &=& \int_{\Sigma} d^3y \sqrt{-h} \frac{1}{8\pi G}\Big( h_{ab} \Delta K - \Delta K_{ab} - 8\pi G \big( t_{ab} + Y_{ab}\big)\, \Big) = 0, 
\end{eqnarray}
where the notation $\Delta A \equiv A_+ - A_-$ means the difference of the values between boundaries of the manifold $\mathcal{M}_+$ and manifold $\mathcal{M}_-$,
 respectively. Finally, the equation of motion on the thin-shell at the boundaries takes form
\begin{eqnarray}
	h_{b}^a \Delta K - \Delta K_{b}^a = 8 \pi G S_b^a ,
		\label{eom_thin_shell}
\end{eqnarray}
where the new effective energy momentum tensor $S_{ab}$ on the thin-shell is defined by,
\begin{eqnarray}
	S_{b}^a &\equiv& = t_{b}^a + Y_{b}^a\,. 
\end{eqnarray}
Furthermore, it is very convenient to represent the $S_b^a$ tensor in the matrix form
\begin{eqnarray}
	S_{b}^a	&\equiv& \left(
\begin{array}{ccc}
-\rho_{\rm eff.}\,~0~\,~0~\\
~\,\, 0~  \,P_{\rm eff.}\,~ 0~ \\
~\,\,\; 0~ \,~ 0~ \,P_{\rm eff.}
\end{array}
\right)
=\left(
\begin{array}{ccc}
-\sigma-\rho_g\,~\quad 0\quad~\,~\quad 0\quad~\\
~\,\, \quad 0\quad~  \,p + p_g^{(\perp)} \quad\; 0\quad~ \\
~\,\,\; \quad 0\quad~ \,~ \quad 0\quad ~p + p_g^{(\perp)}
\end{array}
\right),
\label{S_ab}
\end{eqnarray}
where the explicit forms of the $\rho_g$ and $p_g^{(\perp)}= p_g^{(\theta,\phi)}$ are given in Eqs.(\ref{rho-g},\ref{p-g-theta}). We will see in the latter that the equation of motion of the dRGT massive gravity wormholes takes very simple form like the standard GR case with two types of fluids.
\subsection{The thin-shell wormhole dynamics in dRGT spacetimes}
\label{conditions_dRGT_wormholes}
In this subsection, we consider the two different spacetimes where the thin-shell wormhole connects them together. Thus, the stability of the wormhole can be described by the dynamic of the throat of the wormhole $a(\tau)$ at thin-shell. The junction condition of the thin-shell plays crucial role for describing the wormhole dynamics. Before moving forward to calculate all relevant quantities in the junction condition, we shall summarize some conditions for the thin-shell wormhole spacetime. The first condition is the continuity of the metric tensor condition on thin-shell. This means that the metric tensor of both manifolds are continuous at the throat \cite{Ovgun:2016qiu, Kokubu:2017}
\begin{equation}
g_{\mu\nu}^{+} = g_{\mu\nu}^{-}.
\label{metric_continuous}
\end{equation}
With this condition, the geodesic of particle traveling between two manifolds exists. The second condition is the positiveness condition of metric tensor on thin-shell, i.e., the element of metric tensor must be positive \cite{Ovgun:2016qiu, Kokubu:2017}
\begin{equation}
g_{\mu\nu}^{\pm} > 0.
\end{equation}
This also implies that we are interested in the range of $r$ satisfying this condition, $f_\pm(r) > 0$. 

We next consider the relation between the thin-shell wormhole on the hypersurface $\Sigma$ and the spacetime on manifold $\mathcal{M}_{\pm}$. By matching terms between the metric of the dRGT universe (bulk) and the metric on the thin-shell (hypersurface). Looking at the parameterization of the coordinate on the hypersurface in Eq.(\ref{Hypersur_thin_shell}), we find
\begin{equation}
-d\tau^2 = -f_\pm(a) dT_\pm^2 + \frac{da^2}{f_\pm(a)},
\label{matching_throat}
\end{equation}
and 
\begin{equation}
a^2 d\Omega^2 = r^2 d\Omega^2 ;\quad d\Omega^2 = d\theta^2 + {\rm sin}d\phi^2.
\end{equation}
The relation between $T$ and $\tau$ is given by
\begin{equation}
\dot{T}_\pm = \frac{1}{f_\pm(a)}\sqrt{f_\pm(a) + \dot{a}^2},
\label{tdot}
\end{equation}
where a dot denotes derivative with respect to the proper time, $\tau$, and
\begin{equation}
\ddot{T}_\pm = -\frac{\dot{f}_\pm}{f_\pm^2}\sqrt{f_\pm+\dot{a}^2} + \frac{2 \dot{a} \ddot{a} + \dot{f}_\pm}{2 f_\pm \sqrt{f_\pm + \dot{a}^2 }},
\label{tdotdot}
\end{equation}
where $\dot{f} = \frac{df}{d\tau} = \frac{df(a)}{da} \frac{da}{d\tau} = f'(a) \,\dot{a}$ and a prime denotes the first derivative with respect to $a$. Now we are ready to compute the non-vanished components of the extrinsic curvature of the wormhole in dRGT massive gravity. Having used the line element in Eq.(\ref{metric_Thin_dhell}), one obtains non-vanished components of $K_b^a$ as
\begin{eqnarray}
K^{\tau \pm}_{\tau} &=& \pm \frac{1}{\sqrt{f_{\pm} + \dot{a}^2 }} \bigg( \ddot{a} + \frac{f'_{\pm}}{2} \bigg), \label{k1} \\
K^{\theta \pm}_{\theta} &=& K^{\phi \pm}_{\phi} = \pm \frac{1}{a} \big(\sqrt{ f_{\pm} + \dot{a}^2} \big), \label{k2}
\end{eqnarray}
We note that the extrinsic curvature is a diagonal matrix.
%and the non-zero components of $h^{ac}$ can be calculated from the inverse of $h_{ab}$ in Eq.(\ref{metric_Thin_dhell}) to yield
%\begin{eqnarray}
%	h^{\tau \tau} = -1, \qquad h^{\theta \theta} = h^{\phi \phi} = %\frac{1}{a^2}.
%\end{eqnarray}
It is worth mentioning about the parameters in the dRGT massive gravity. By using the observational constraints on the dRGT theory, we find
\begin{eqnarray}
\zeta = 0,\quad\Rightarrow\quad \alpha = -3\beta = -\frac{3}{2} - \frac{\Lambda}{2 m_g^2},\quad k = \frac{2\gamma}{m_g^2+\Lambda}\,.
\end{eqnarray}
By using Eqs.(\ref{rho-g},\ref{p-g-theta}) with $r=a$ at the boundaries $\partial\mathcal{M}_\pm$, the components of the effective energy momentum tensor, $S_b^a$, in Eq.(\ref{S_ab}) are given by the following explicit expressions,
\begin{eqnarray}
S_\tau^\tau &=& -\rho_{\rm eff.} = -\sigma - \rho_g(a)
\nonumber\\
&=& -\sigma + \frac{1}{8\pi G}\left( \frac{2\gamma}{a} - \Lambda\right), 
\nonumber\\
S_\theta^\theta &=& S_\phi^\phi = P_{\rm eff.} = p + p_g^{(\perp)}(a)
\nonumber\\
&=& p + \frac{1}{8\pi G}\left( \frac{\gamma}{a} - \Lambda\right).
\end{eqnarray}
According to the $\zeta = 0$ constraint, we remarkably found that there is only one free parameter of the dRGT massive gravity since the graviton mass, $m_g$ and the cosmological constant, $\Lambda$ can be fixed by using the observed values of those two quantities.

In this work, we employ the $Z_2$ symmetry between two metric tensors of the manifolds. This means that $f_+(a) = f_-(a) = f(a)$. The $(\tau\tau)$ component of the junction condition of the thin-shell wormhole in Eq.(\ref{eom_thin_shell}) reads
\begin{eqnarray}
	\frac{2}{a} \big(\sqrt{ f + \dot{a}^2} \big) = -8\pi G \sigma + \left( \frac{2\gamma}{a} - \Lambda\right).
	\label{energy_density_thin_shell}
\end{eqnarray}
On the other hand, the angular component of Eq.(\ref{eom_thin_shell}) is given by
\begin{equation}
	\frac{1}{\sqrt{f + \dot{a}^2 }} \big( 2\ddot{a} + f' \big) = 8\pi G p + \left( \frac{\gamma}{a} - \Lambda\right).
	\label{pressure_density_thin_shell}
\end{equation}
In addition, the continuity of the perfect fluid matter gives the relation between the energy density and pressure on the thin-shell wormhole as
\begin{equation}
	\frac{d}{d \tau}\big( a \sigma \big) + p \frac{da}{d \tau} = 0.
	\label{energy_pressure_relation}
\end{equation}
It is also written in terms of the first order derivative of $\sigma$ with respect to $a$ as
\begin{equation}
	\frac{d \sigma}{d a} = -\bigg( \frac{\sigma + p}{a} \bigg).
	\label{first_order_sigma}
\end{equation}
The second order derivative of $\sigma$ with respect to $a$ yields
\begin{equation}
	\frac{d^2 \sigma}{da^2} = \frac{\sigma + p}{a^2} \bigg( 2 + \frac{dp}{d \rho} \bigg),
	\label{second_order_sigma}
\end{equation}
where $p = p(\sigma)$. Above equations will be useful for investigating the stability of the wormhole with several types of the perfect fluid matters. We study their effects on particular models in the next section.

%%%%%%%%%%%%%%%%%%%%%%%%%%%%%%%%%%%%%%%%%%
\section{Stability analysis of the dRGT thin-shell wormhole}\label{stability}
%%%%%%%%%%%%%%%%%%%%%%%%%%%%%%%%%%%%%%%%
The stability of the wormhole can be quantified by the study of the effective potential of the wormhole dynamics. The equation of motion for determining the stability of the throat $a(\tau)$ is directly derived from Eq.(\ref{energy_density_thin_shell}) to obtain 
\begin{equation}
	\frac12 \dot{a}^2 + V(a) = 0,
	\label{master_equation_throat}
\end{equation}
where the effective potential $V(a)$ is written by
\begin{equation}
	V(a) = \frac12f(a) - \frac{a^2}{8} \left[ 8\pi G \sigma - \left( \frac{2\gamma}{a} - \Lambda\right) \right]^2.
	\label{va}
\end{equation}
This single dynamical equation (\ref{master_equation_throat}) completely determines
the motion of the wormhole throat. Notice that if we consider only the massive gravity correction term, $m^{2}_{g}{\cal U}$, without invoking exotic fluids, i.e., $\sigma=0=p$, the potential (\ref{va}) becomes
\begin{equation}
	V(a) = \frac12f(a) - \frac{a^2}{8} \left[  \frac{2\gamma}{a} - \Lambda \right]^2. \label{vaa}
\end{equation}
In this situation, we find that $V''(a_0)<0$ meaning that it is not possible to obtain stable wormholes if $\Lambda>0$. We assume that the throat of the thin-shell wormhole is static at $a = a_0$ and satisfies the relation
\begin{equation}
f(a_0) > 0,
\label{cdt_positiven}
\end{equation}
in order to avoid the event horizon $r_{\text{EH}}$ from the wormhole, $f(r_{\text{EH}}) = 0$.
In order to analyze the stability of the throat, we employ a small perturbation to the potential and are able to determine whether the throat is stable or not. Here the usual Taylor series expansion is applied to the potential $V(a)$ around the static radius $a_0$ as follows:
\begin{equation}
V(a) = V(a_0) + V'(a_0) (a - a_0)+ \frac{1}{2} V''(a_0)(a - a_0)^2 + \mathcal{O}((a - a_0)^3).
\label{Taylor_potential}
\end{equation}
When evaluating at the static solution $a=a_{0}$, we obtain the expected result $V(a_0) = 0$ and $ V'(a_0) = 0$ if $a_0$ is the static radius. Then, the Eq.~(\ref{Taylor_potential}) reduces to
\begin{equation}
V(a) = \frac{1}{2} V''(a_0)(a - a_0)^2 + \mathcal{O}((a - a_0)^3).
\label{Taylor_potential2}
\end{equation}
Therefore, the equation of motion for the wormhole throat approximately takes the form 
\begin{equation}
\dot{a}^2 + \frac{1}{2} V''(a_0)(a - a_0)^2 = 0.
\label{master_throat_perturb}
\end{equation}
Using Eq.(\ref{va}) with the help of  Eq.~(\ref{first_order_sigma}) and Eq.~(\ref{second_order_sigma}), we find
\begin{eqnarray}
	V''(a_0) = \frac{1}{2} f''(a_0) &+& \frac{d p}{d \sigma} \bigg( -2 G (p + \sigma) \pi \Lambda - 16 G^2 \pi^2 \sigma (p + \sigma) + \frac{4 G  \pi \gamma (p + \sigma)}{a} \bigg) \nonumber \\
	&-& 16 G^2 p^2 \pi^2 + 4 G p \pi \Lambda - \frac{1}{4} \Lambda^2.
	\label{vprimeprime}
\end{eqnarray}
Thus, the wormhole is stable if and only if $V''(a_0)>0$  where the motion of the throat is oscillatory with angular frequency $\omega=\sqrt{V''(a_{0})/2}$. Note that $V(a_{0})$ has a local minimum at $a_{0}$. To carry out the analysis, we can quantify which conditions we obtain stable wormholes. In our case, we find that these parameters need to satisfy
\begin{eqnarray}
	0 < \frac{1}{2} f''(a_0) &+& \frac{d p}{d \sigma} \bigg( -2 G (p + \sigma) \pi \Lambda - 16 G^2 \pi^2 \sigma (p + \sigma) + \frac{4 G \pi \gamma (p + \sigma)}{a} \bigg) \nonumber \\
	&-& 16 G^2 p^2 \pi^2 + 4 G p \pi \Lambda - \frac{1}{4} \Lambda^2.
	\label{va0}
\end{eqnarray}
Next we employ four fluid models for studying the stability of the dRGT wormhole: (1) a linear model, (2) a Chaplygin gas model, (3) a generalized Chaplygin gas model and (4) a logarithm model. It has been shown in the previous section that we have only one free parameter ($\gamma$) in the dRGT theory by using $\zeta = 0$. The remained parameters in this work in the natural unit are given by
\begin{eqnarray}
    G &=& 6.72 \times 10^{-57} ~{\rm eV}^{-2},\quad m_g = 1.22 \times 10^{-22} ~{\rm eV},\quad  \Lambda = 4.33 \times 10^{-66}~ {\rm eV}^2, 
\end{eqnarray}
where the gravitational constant $G$, the cosmological constant, $\Lambda$ are taken from review of particle physics \cite{Tanabashi:2018oca} and the graviton mass is the upper bound values from the LIGO-VIRGO gravitationa wave observations \cite{Abbott:2016blz}. Moreover, the free parameter, $\gamma$ of the dRGT massive gravity, has been fixed by fitting rotational curves of the galaxies in several data \cite{Panpanich:2018cxo} and we will use this value for the following study. The $\gamma$ parameter and the thin-shell wormhole mass read
\begin{eqnarray}
    \gamma &=& 6.05 \times 10^{-34} ~{\rm eV},\quad M = 3.36 \times 10^{66} ~{\rm eV},
\end{eqnarray}
where we have assumed that $M$ is roughly equal to the lower bound mass of black hole which is three times of the solar mass. Using the above numerical values, we can estimate the event horizon and the cosmological horizon via Eq.(\ref{fr_dRGT}) by setting $f(r)=0$.

However, it is much more convenient to use the dimensionless values of physical parameters given below:
\begin{eqnarray}
     \quad \Lambda = 0.0001,\quad  \gamma = 0.001,\quad  M = 1.
\end{eqnarray}
Note that one can easily change the units from this dimensionless parameters to the others, e.g. SI units or Natural units, and vice versa. Now we consider the energy density given in Eq.~(\ref{energy_density_thin_shell}) and write for a static case at the throat as:
\begin{eqnarray}
	\sigma = \frac{1}{8\pi G} \left( \frac{2\gamma}{a} - \Lambda\right) - \frac{2}{8\pi G a} \sqrt{f}. 
	\label{energy_density}
\end{eqnarray}
%The behaviors of the energy density $\sigma(a)$ from the  is illustrated in the range between the event and cosmological horizon in the Fig.\ref{figure_sigma}. 
Here we can solve the above equation to write $\sigma$ in terms of $a$ when substituting a function $f(a)$. It was found that the stability of transparent spherically symmetric thin shells to linearized spherically symmetric perturbations about static equilibrium has been examined Ref.\cite{Ishak:2001az}.

\subsection{Linear model}
\begin{figure}[h]
	\includegraphics[width=8cm]{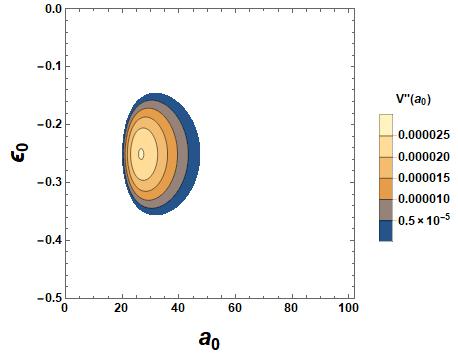}
	\centering
	\caption{The plot shows the stable region of the linear model $p(\sigma) = \epsilon_0 \sigma$. The contour shows that the constant $\epsilon_{0}$ has negative values in the throat.}
	\label{figure_stable_linear}
\end{figure}

We begin our stability analyses by considering the pressure which is proportional the energy density \cite{Ovgun:2016qiu}:
\begin{equation}
p(\sigma) = \epsilon_0 \sigma.
    \label{p_linear}
    \end{equation}
It is easy to show that
\begin{equation}
    \frac{d p}{d \sigma} = \epsilon_0.
    \label{d_p_linear}
\end{equation}
Notice that the change in the pressure on the energy density is a constant. Moreover, the throat of the traversable wormhole basically locates between the event horizon and the cosmological horizon. After substituting the above results into the stability condition (\ref{va0}), we find
\begin{eqnarray}
	0 < \frac{1}{2} f''(a_0) &-& \frac{1}{4} \bigg(  \Lambda^2 + 8 \pi ( \epsilon_0 - 1) \epsilon_0 \Lambda \sigma + 64 \pi^2 \epsilon_0 (1 + 2 \epsilon_0) \sigma^2  \bigg) \nonumber \\
	&+& \frac{4 \pi \gamma \epsilon_0 (\epsilon_0 + 1)}{a} .
	\label{va0_linear}
\end{eqnarray}
In order to visualize the stability region of the model, we plot the stability contour in terms of $\epsilon_{0}$ and $a_{0}$. Our result is illustrated in Fig.\ref{figure_stable_linear} for the linear model. 
We notice that in order to satisfy the stability condition (\ref{va0}) the constant $\epsilon_{0}$ has negative values in the throat radius $a_{0}$. 

\subsection{Chaplygin gas model}
\begin{figure}[h]
	\includegraphics[width=8cm]{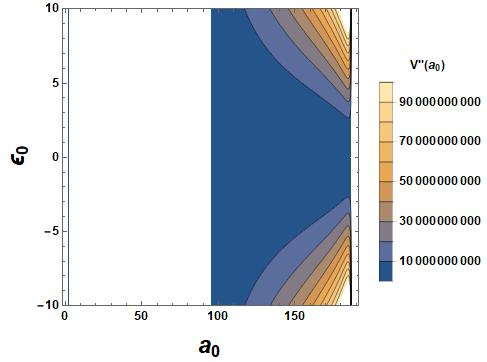}
	\centering
	\caption{The plot shows the stable region of the Chaplygin gas model $p(\sigma) = -\epsilon_0 ( \frac{1}{\sigma} - 	\frac{1}{\sigma_0})$. The result shows that $\epsilon_{0}$ can have both negative values and positive ones in the throat with radius $a_{0}$.}
	\label{figure_stable_Chap}
\end{figure}

We next consider the Chaplygin gas model for the exotic matter. The pressure is already given in Ref.\cite{Ovgun:2016qiu}:
\begin{equation}
	p(\sigma) = -\epsilon_0 \bigg( \frac{1}{\sigma} - 	\frac{1}{\sigma_0} \bigg).
	\label{p_Chap}
\end{equation}
It is trivial to show that
\begin{equation}
	\frac{d p}{d \sigma} = \frac{\epsilon_0}{\sigma_0^2},
	\label{d_p_Chap}
\end{equation}
where $\sigma_0$ is a constant. After substituting the above results into the stability condition (\ref{va0}), we find in this case
\begin{eqnarray}
    0 < \frac{1}{2} f''(a_0) &+& \frac{1}{4 a \sigma^3 \sigma_0^2} \bigg( 16 \pi \gamma \epsilon_0 \sigma_0 (\epsilon_0(\sigma - \sigma_0) + \sigma^2 \sigma_0) - a \big( \Lambda^2 \sigma^3 \sigma_0^2 \nonumber \\
    &+& 64 \pi^2 \epsilon_0 \sigma^2 (\epsilon_0 (\sigma - \sigma_0) + \sigma \sigma_0^2 ) + 8  \pi  \epsilon_0 \Lambda \sigma_0 ( \epsilon_0 (\sigma - \sigma_0) + \sigma^2 (3 \sigma_0 - 2 \sigma) )  \big) \bigg). \\
\end{eqnarray}
Here we plot the stability contour in terms of $\epsilon_{0}$ and $a_{0}$ for this model. The result is illustrated in Fig.\ref{figure_stable_Chap}. The stable region for this case is represented in Fig.\ref{figure_stable_Chap}.
We notice that in order to satisfy the stability condition (\ref{va0}) $\epsilon_{0}$ can have both negative values and positive ones in the throat with radius $a_{0}$.

\subsection{Generalized Chaplygin gas model}
\begin{figure}[h]
	\includegraphics[width=8cm]{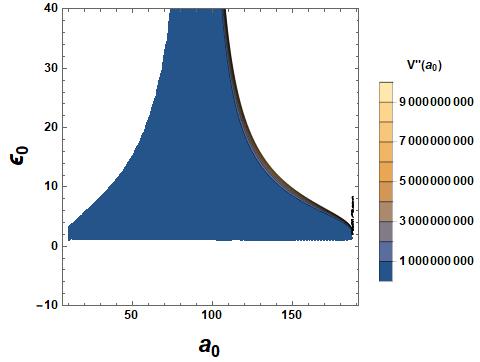}
	\centering
	\caption{The plot shows the stable region of the generalized Chaplygin gas model $p(\sigma) = ( \frac{\sigma_0}{\sigma})^{\epsilon_0}$. The result shows that $\epsilon_{0}$ has positive values in the throat with $a_{0}$.}
	\label{figure_stable_gen_Chap}
\end{figure}

In addition, the Chaplygin gas model given in the previous subsection can be generalized where the relation between $p(\sigma)$ and $\sigma$ takes the form \cite{Ovgun:2016qiu}
\begin{equation}
    p(\sigma) = \bigg( \frac{\sigma_0}{\sigma} \bigg)^{\epsilon_0},
    \label{p_gen_Chap}
\end{equation}
and
\begin{equation}
    \frac{d p}{d \sigma} = - \epsilon_0 \frac{\sigma_0^{\epsilon_0}}{\sigma^{\epsilon_0 + 1}}.
    \label{d_p_gen_Chap}
\end{equation}
After substituting the above results into the stability condition (\ref{va0}), we find in this case
\begin{eqnarray}
    0 < \frac{1}{2} f''(a_0) &+& \frac{1}{4a} \sigma^{-1-\epsilon_0} \bigg( -64 a \pi^2 \bigg( \frac{\sigma_0^{2 \epsilon_0}}{\sigma^{\epsilon_0 - 1}} \bigg) + 8 \pi \bigg( \frac{\sigma_0}{\sigma}  \bigg)^{\epsilon_0} (-2 a \Lambda \sigma^{1 + \epsilon_0} + \epsilon_0 ( (-2 \gamma + a \Lambda) \nonumber \\
    &+& 8 a \pi \sigma) \sigma_0^{\epsilon_0}) + \sigma \bigg(- 8 \pi \epsilon_0 ((-2 \gamma + a \Lambda ) + 8 a \pi \sigma ) \sigma_0^{\epsilon_0}\bigg) \bigg) - \frac{1}{4} \Lambda^2.
	\label{va0_gen_Chap}
\end{eqnarray}
Here we display the stability contour in terms of $\epsilon_{0}$ and $a_{0}$ illustrated in Fig.\ref{figure_stable_gen_Chap}. The stable region for this case is represented in Fig.\ref{figure_stable_gen_Chap} for the generalized Chaplygin gas model. 
We find that in order to satisfy the stability condition (\ref{va0}) $\epsilon_{0}$ has positive values in the throat with radius $a_{0}$.

\subsection{Logarithm model}
\begin{figure}[h]
	\includegraphics[width=8cm]{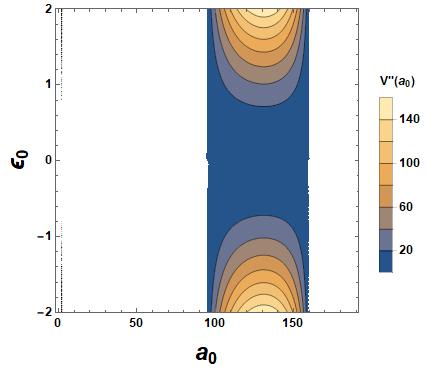}
	\centering
	\caption{The plot shows the stable region of the linear model $p(\sigma) = \epsilon_0 \text{log}( \frac{\sigma}{\sigma_0})$. The result shows that $\epsilon_{0}$ can have both negative values and positive ones in the throat with radius $a_{0}$.}
	\label{figure_stable_log}
\end{figure}

We provide the last example in which the pressure $p(\sigma)$ and the energy density $\sigma$ are related via \cite{Ovgun:2016qiu}
\begin{equation}
    p(\sigma) = \epsilon_0 \text{log}\bigg( \frac{\sigma}{\sigma_0} \bigg),
    \label{p_log}
\end{equation}
and
\begin{equation}
    \frac{d p}{d \sigma} = \frac{\epsilon_0}{\sigma}.
    \label{d_p_log}
\end{equation}
After substituting the above results into the stability condition (\ref{va0}), we find in this particular case
\begin{eqnarray}
    0 < \frac{1}{2} f''(a_0) &+& \frac{4 \pi \gamma \epsilon_0 \bigg(\sigma + \epsilon_0 \text{log}\big( \frac{\sigma}{\sigma_0} \big) \bigg) }{\sigma a} - \frac{ \Lambda^2 + 8 \pi \Lambda \big( \epsilon_0 + \epsilon_0 (\frac{\epsilon_0}{\sigma} - 2) \text{log} \big( \frac{\sigma}{\sigma_0} \big) \big) }{4 a}\nonumber \\
    &-& \frac{64 \pi^2 \epsilon_0 \bigg( \sigma + \epsilon_0 \text{log} \big( \frac{\sigma}{\sigma_0} \big)\big( 1 + \text{log} \big( \frac{\sigma}{\sigma_0} \big) \big) \bigg) }{4 a}.
\end{eqnarray}
Here we display the stability contour in terms of $\epsilon_{0}$ and $a_{0}$ illustrated in Fig.\ref{figure_stable_log}. The stable region for this case is represented in Fig.\ref{figure_stable_log} for the logarithm model model.
We observe that in order to satisfy the stability condition (\ref{va0}) $\epsilon_{0}$ can have both negative values and positive ones in the throat with radius $a_{o}$.

%%%%%%%%%%%%%%%%%%%%%%%%%%%%%
\section{Energy Conditions}
\label{ener}
%%%%%%%%%%%%%%%%%%%%%%%%%%%%%
In this section, we shall analyze the energy conditions for the thin-shell wormholes in the dRGT massive gravity. We check the null, weak, and strong conditions at the wormhole throat for all existing models present in the previous section. 

\begin{itemize}

\item[I.]{Null energy condition is expressed in terms of energy density and pressure as follows:
\begin{equation}
\rho_{\rm eff.}+P_{\rm eff.} \geq 0,
\end{equation}
which yields
\begin{eqnarray}
\rho_{\rm eff.}+P_{\rm eff.} &=& \sigma - \frac{1}{8\pi G}\left( \frac{2\gamma}{a} - \Lambda\right) + p + \frac{1}{8\pi G}\left( \frac{\gamma}{a} - \Lambda\right)\nonumber\\&=& \sigma+p - \frac{1}{8\pi G}\frac{\gamma}{a}\geq 0.
\label{condt1}
\end{eqnarray}}

\item[II.]{Weak energy condition is given by
\begin{equation}
\rho_{\rm eff.}\geq 0,\,\,\,\,\rho_{\rm eff.}+P_{\rm eff.} \geq 0,
\end{equation}
which gives the following result for the thin-shell wormholes in the dRGT massive gravity
\begin{equation}
\rho_{\rm eff.}=\sigma - \frac{1}{8\pi G}\left( \frac{2\gamma}{a} - \Lambda\right)\geq 0,
\label{condt2}
\end{equation}}

\item[III.]{Strong energy condition is governed by
\begin{equation}
\rho_{\rm eff.} + 3P_{\rm eff.} \geq 0,\,\,\,\,\rho_{\rm eff.}+P_{\rm eff.} \geq 0,
\end{equation}
which gives the following result for the thin-shell wormholes in the dRGT massive gravity
\begin{eqnarray}
\rho_{\rm eff.}+3P_{\rm eff.} &=& \sigma - \frac{1}{8\pi G}\left( \frac{2\gamma}{a} - \Lambda\right) + 3p + \frac{3}{8\pi G}\left( \frac{\gamma}{a} - \Lambda\right)\nonumber\\&=&\sigma+3p + \frac{1}{8\pi G}\left( \frac{\gamma}{a} - 2\Lambda\right)\geq 0.
\label{condt3}
\end{eqnarray}}

\end{itemize}

\subsection{Linear model}
When substituting the pressure and the energy density of this model into Eq.~(\ref{condt1}), Eq.~(\ref{condt2}) and Eq.~(\ref{condt3}), we find
\begin{eqnarray}
    \rho_{\rm eff.} + P_{\rm eff.} &=& \frac{( (1 + 2 \epsilon_0)\gamma - a (1 + \epsilon_0) \Lambda ) - 2 (1 + \epsilon_0) \sqrt{f(a)} }{8 \pi G a} \geq 0, \\
    \rho_{\rm eff.} &=& -\frac{\sqrt{f(a)}}{4 \pi G a} \geq 0, \\
    \rho_{\rm eff.} + 3P_{\rm eff.} &=& \frac{3 ( (1 + 2 \epsilon_0)\gamma - a (1 + \epsilon_0) \Lambda ) - 2 (1 + \epsilon_0) \sqrt{f(a)} }{8 \pi G a} \geq 0.
\end{eqnarray}
\begin{figure}[h]
	\includegraphics[width=7.2cm]{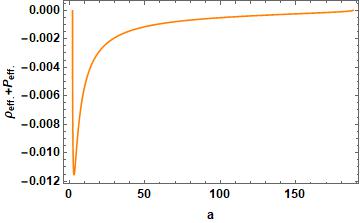}
	\includegraphics[width=7.4cm]{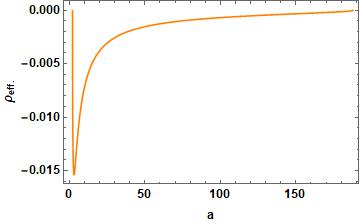}
	\includegraphics[width=7.4cm]{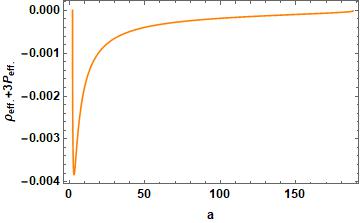}
	\centering
	\caption{The plots show the variation of $\rho_{\rm eff.} + P_{\rm eff.},\,\rho_{\rm eff.}$ and $\rho_{\rm eff.} + 3P_{\rm eff.}$ as a function of $a$ of the linear model with $p(\sigma) = \epsilon_0 \sigma$.} 
	\label{ECLinear}
\end{figure}
In order to analyse the energy conditions, we will choose the values of $\epsilon_{0}$ in the stable regions shown in Fig.\ref{figure_stable_linear} and then verify the energy conditions. Fig.\ref{ECLinear} shows the variation of $\rho_{\rm eff.} + P_{\rm eff.},\,\rho_{\rm eff.}$ and $\rho_{\rm eff.} + 3P_{\rm eff.}$ as a function of $a$ in the linear model $p(\sigma) = \epsilon_0 \sigma$. We observe that all energy conditions are violated in this model.

\subsection{Chaplygin gas model}
We substitute the pressure and the energy density of this model into Eq.~(\ref{condt1}), Eq.~(\ref{condt2}) and Eq.~(\ref{condt3}) and then we obtain
\begin{eqnarray}
    \rho_{\rm eff.} + P_{\rm eff.} &=& \frac{1}{8}\bigg( \frac{(\gamma - a \Lambda)}{\pi G a} - \frac{8 \epsilon_0 ( (-2\gamma + a \Lambda) + 8 a G \pi \sigma_0 + 2 \sqrt{f(a)} )}{\sigma_0 ((2 \gamma - a \Lambda) - 2 \sqrt{f(a)} )} - \frac{2 \sqrt{f(a)}}{a G \pi} \bigg) \geq 0,\nonumber \\
    \rho_{\rm eff.} &=& -\frac{\sqrt{f(a)}}{4 \pi G a} \geq 0, \\
    \rho_{\rm eff.} + 3P_{\rm eff.} &=& \frac{1}{8}\bigg( \frac{3 (\gamma - a \Lambda)}{\pi G a} - \frac{24 \epsilon_0 ( (-2\gamma + a \Lambda) + 8 a G \pi \sigma_0 + 2 \sqrt{f(a)} )}{\sigma_0 ((2 \gamma - a \Lambda) - 2 \sqrt{f(a)} )} - \frac{2 \sqrt{f(a)}}{a G \pi} \bigg) \geq 0. \nonumber \\
\end{eqnarray}
\begin{figure}[h]
	\includegraphics[width=7.2cm]{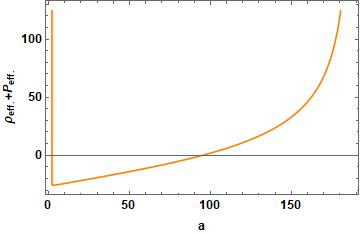}
	\includegraphics[width=7.4cm]{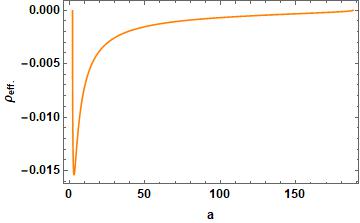}
	\includegraphics[width=7.4cm]{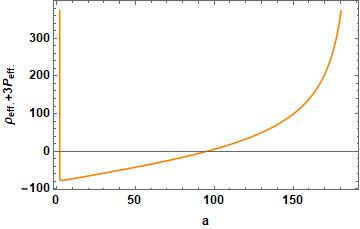}
	\centering
	\caption{The plots show the variation of $\rho_{\rm eff.} + P_{\rm eff.},\,\rho_{\rm eff.}$ and $\rho_{\rm eff.} + 3P_{\rm eff.}$ as a function of $a$ of the Chaplygin gas model with $p(\sigma) = -\epsilon_0 ( \frac{1}{\sigma} - 	\frac{1}{\sigma_0})$.}
	\label{ECCgas}
\end{figure}
In order to quantify the energy conditions, we will choose the values of $\epsilon_{0}$ in the stable regions shown in Fig.\ref{figure_stable_Chap} and then examine the energy conditions. Fig.\ref{ECCgas} shows the variation of $\rho_{\rm eff.} + P_{\rm eff.},\,\rho_{\rm eff.}$ and $\rho_{\rm eff.} + 3P_{\rm eff.}$ as a function of $a$ in the linear model $p(\sigma) = -\epsilon_0 ( \frac{1}{\sigma} - 	\frac{1}{\sigma_0})$. We observe that all energy conditions are violated for $a<100$ in this model.

\subsection{Generalized Chaplygin gas model}
When substituting the pressure and the energy density of this model into Eq.~(\ref{condt1}), Eq.~(\ref{condt2}) and Eq.~(\ref{condt3}), we find
\begin{eqnarray}
    \rho_{\rm eff.} + P_{\rm eff.} &=& \frac{1}{8 \pi G a} \bigg( (\gamma - a \Lambda) - 2 \sqrt{f(a)} \nonumber \\
    &-& G a (8 \pi)^{1 + \epsilon_0} \bigg( - \frac{G a \sigma_0}{(-2\gamma + a \Lambda) + 2 \sqrt{f(a)}} \bigg)^{\epsilon_0} \bigg) \geq 0, \\
    \rho_{\rm eff.} &=& -\frac{\sqrt{f(a)}}{4 \pi G a} \geq 0, \\
    \rho_{\rm eff.} + 3P_{\rm eff.} &=& \frac{1}{8 \pi G a} \bigg( 3 (\gamma - 3 a \Lambda) - 2 \sqrt{f(a)} \nonumber \\
    &-& 3 G a (8 \pi)^{1 + \epsilon_0} \bigg( - \frac{G a \sigma_0}{(-2\gamma + a \Lambda) + 2 \sqrt{f(a)}} \bigg)^{\epsilon_0}  \bigg) \geq 0.
\end{eqnarray}
\begin{figure}[h]
	\includegraphics[width=7.2cm]{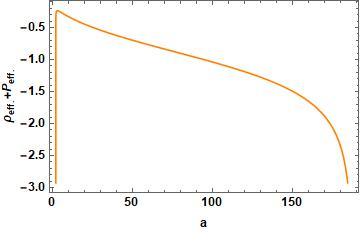}
	\includegraphics[width=7.4cm]{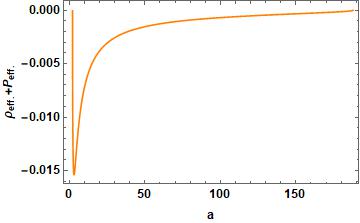}
	\includegraphics[width=7.4cm]{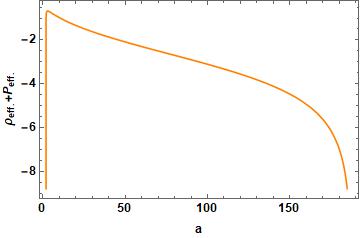}
	\centering
	\caption{The plots show the variation of $\rho_{\rm eff.} + P_{\rm eff.},\,\rho_{\rm eff.}$ and $\rho_{\rm eff.} + 3P_{\rm eff.}$ as a function of $a$ of the generalized Chaplygin gas model with $p(\sigma) = ( \frac{\sigma_0}{\sigma})^{\epsilon_0}$.}
	\label{ECgCgas}
\end{figure}
We here quantify the energy conditions by choosing the values of $\epsilon_{0}$ in the stable regions shown in Fig.\ref{figure_stable_gen_Chap} and then examine the energy conditions. Fig.\ref{ECgCgas} shows the variation of $\rho_{\rm eff.} + P_{\rm eff.},\,\rho_{\rm eff.}$ and $\rho_{\rm eff.} + 3P_{\rm eff.}$ as a function of $a$ in the generalized Chaplygin gas model $p(\sigma) = (\frac{\sigma_0}{\sigma})^{\epsilon_0}$. We observe that all energy conditions are violated for positive values of $\epsilon_{0}$.

\subsection{Logarithm model}
We substitute the pressure and the energy density of this model into Eq.~(\ref{condt1}), Eq.~(\ref{condt2}) and Eq.~(\ref{condt3}) and then we obtain
\begin{eqnarray}
\rho_{\rm eff.} + P_{\rm eff.} &=& \frac{(\gamma - a \Lambda) - 2 \sqrt{f(a)}}{8 \pi G a}+ \epsilon_0 \text{log} \bigg( \frac{(2 \gamma - a \Lambda) - 2 \sqrt{f(a)}}{8 \pi G a \sigma_0} \bigg) \geq 0,\\
    \rho_{\rm eff.} &=& -\frac{\sqrt{f(a)}}{4 \pi G a} \geq 0,\\
    \rho_{\rm eff.} + 3P_{\rm eff.} &=& \frac{3(\gamma - a \Lambda) - 2 \sqrt{f(a)}}{8 \pi G a} + 3 \epsilon_0 \text{log} \bigg(\frac{(2 \gamma - a \Lambda) - 2 \sqrt{f(a)}}{8 \pi G a \sigma_0} \bigg) \geq 0.
\end{eqnarray}
\begin{figure}[h]
	\includegraphics[width=7.2cm]{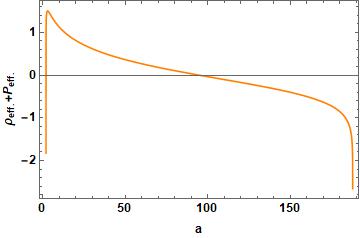}
	\includegraphics[width=7.4cm]{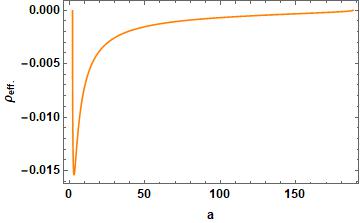}
	\includegraphics[width=7.4cm]{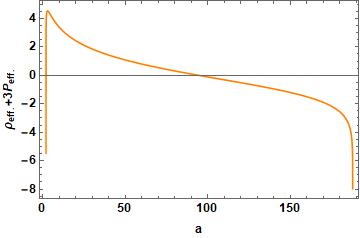}
	\centering
	\caption{The plots show the variation of $\rho_{\rm eff.} + P_{\rm eff.},\,\rho_{\rm eff.}$ and $\rho_{\rm eff.} + 3P_{\rm eff.}$ as a function of $a$ of the log model with $p(\sigma) = \epsilon_0 \text{log}( \frac{\sigma}{\sigma_0})$.}
	\label{ECLog}
\end{figure}
We here quantify the energy conditions by choosing the values of $\epsilon_{0}$ in the stable regions shown in Fig.\ref{figure_stable_log} and then examine the energy conditions. Fig.\ref{ECLog} shows the variation of $\rho_{\rm eff.} + P_{\rm eff.},\,\rho_{\rm eff.}$ and $\rho_{\rm eff.} + 3P_{\rm eff.}$ as a function of $a$ in the ogarithm model $p(\sigma) = \epsilon_0 \text{log}( \frac{\sigma}{\sigma_0})$. We observe that all energy conditions are violated for positive values of $\epsilon_{0}$ with $a>100$.

%%%%%%%%%%%%%%%%%%%%%%%%%%%%%%%%%
\section{Embedding diagram}
\label{dia}
\begin{figure}[h]	
	\includegraphics[width=7cm]{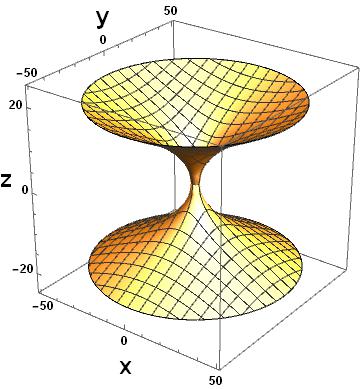}
	\centering
	\caption{The figure shows the wormhole shape obtained via the embedding diagrams in 3d asymptotic flat spacetime. We use $M=1, \Lambda=0.0001, \gamma=0.001$ and set $\zeta=0$.}
	\label{embeds}
\end{figure}

%%%%%%%%%%%%%%%%%%%%%%%%%%
In this section, we construct the wormhole geometry via the embedding diagrams to represent a thin-shell wormhole in the dRGT massive gravity and extract some useful information by considering an equatorial slice, $\theta=\pi/2$ and a fixed moment of time, $t = const.\,$. Therefore the metric reduces to
\begin{equation}
	ds^2 = \frac{dr^2}{f(r)} + r^2 d\phi^2.
	\label{embed}
\end{equation}
where
\begin{eqnarray}
f(r) = 1 - \frac{2M}{r} - \frac{\Lambda r^2}{3} + \gamma r + \zeta.
\label{embedf}
\end{eqnarray}
Next we embed the metric from Eq.(\ref{embedf}) into three-dimensional Euclidean space to visualize this slice and hence the spacetime can be written in cylindrical coordinates as
\begin{equation}
	ds^2 = dz^{2}+dr^{2}+ r^2 d\phi^2=\Big(1+\Big(\frac{dz}{dr}\Big)^{2}\Big)dr^{2}+r^2 d\phi^2.
	\label{embed2}
\end{equation}
Comparing Eq.(\ref{embed}) with Eq.(\ref{embed2}) generates the expression for the embedding surface which is given by
\begin{equation}
	\frac{dz}{dr}= \pm \sqrt{\frac{1-f(r)}{f(r)}}.
	\label{embed12}
\end{equation}
where $f(r)$ is given in Eq.(\ref{embedf}). However, the integration of the above expression can not be solved analytically. Performing an numerical technique allows us to illustrate the wormhole shape given in Fig.\ref{embeds}.

%%%%%%%%%%%%%%%%%%%%%%%%%%%%%%%%%%%
\section{Discussions and conclusions}
\label{discussions}
%%%%%%%%%%%%%%%%%%%%%%%%%%%%%%%
In this work, we have studied the thin-shell wormholes in dRGT massive gravity. In order to construct the thin-shell wormhole, two bulks of the spacetime geometry are glued together via the cut-and-paste procedure \cite{Visser:1989kh}. Moreover, the junction conditions of dRGT spacetime are also derived in this work. The massive graviton correction term of the dRGT theory, $m_g^2\mathcal{U}(g,\phi^2)$, in the Einstein equation is represented in terms of effective anisotropic pressure fluid. However if there is only this correction term, without invoking exotic fluids, we have found that the thin-shell wormholes can not be stabilized. We have also quantified the dynamics of the spherical thin-shell wormholes in the dRGT theory. We then examined the stability conditions of the wormholes by introducing four existing models of the exotic fluids at the throat. In addition, we analyzed the energy conditions for the thin-shell wormholes in the dRGT massive gravity by checking the null, weak, and strong conditions at the wormhole throat.

We have quantified the energy conditions of the four models by choosing the values of $\epsilon_{0}$ in the stable regions shown in Sec.\ref{stability}. We have shown the variation of $\rho_{\rm eff.} + P_{\rm eff.},\,\rho_{\rm eff.}$ and $\rho_{\rm eff.} + 3P_{\rm eff.}$ as a function of $a$ in all models: (1) a linear model $p(\sigma) = (\frac{\sigma_0}{\sigma})^{\epsilon_0}$, (2) a Chaplygin gas model $p(\sigma) = -\epsilon_0 ( \frac{1}{\sigma} -\frac{1}{\sigma_0})$, (3) a generalized Chaplygin gas model $p(\sigma) = ( \frac{\sigma_0}{\sigma})^{\epsilon_0}$ and (4) a logarithm model $p(\sigma) = \epsilon_0 \text{log}( \frac{\sigma}{\sigma_0})$. Using the values of $\epsilon_{0}$ in the stable regions, we have observed that that in general the classical energy conditions are violated by introducing all existing models of the exotic fluids. Moreover, we have quantified the wormhole geometry by using the embedding diagram to represent a thin-shell wormhole in the dRGT massive gravity.

However, there are some limitations in the present work - for example, the construction of the shadow cast by the thin-shell wormhole in the dRGT massive gravity is worth investigating. Regarding this, we can evaluate the test particle geodesics and determine the trajectories of
photons around the wormhole. This can be straightforwardly done by following the work studied in Refs.\cite{Jusufi:2017drg,Ovgun:2018prw,Amir:2018pcu}. Additionally, we can elaborate our work by studying the gravitational lensing effect in the spacetime of the wormhole metric (\ref{fr_dRGT}). This allows us to determine the deflection angle of the photon due to the presence of the wormhole in the dRGT massive gravity. It is worth mentioning that gravitational lensing and particle motions around non-asymptotically flat black hole spacetime in dRGT massive gravity have been done in Ref.\cite{Panpanich:2019mll}.

\acknowledgments

TT thanks the Science Achievement Scholarship of Thailand (SAST) for financial support during his PhD study. AC is supported by the CUniverse research promotion project of Chulalongkorn University under the grant reference CUAASC. DS is supported by Thailand Research Fund (TRF) under a contract No.TRG6180014. PC thanks Kimet Jusufi for initiating research collaboration on the wormhole issues.

%%%%%%%%%%%%%%%%%%%%%%%%%%%%%%%%%%%%%%%%
%%%%%%%%%%%%%%%%%%%%%%%%%%%%%%%%%%%%%%%%
\end{document}